\documentclass[a4paper,twoside]{article}
\newcommand{\affil}[1]{$^{\rm #1}$}
\usepackage[authoryear]{natbib}
\bibpunct{(}{)}{;}{a}{}{,}
\usepackage{graphicx}
\usepackage{url}
\usepackage{fixltx2e}
\date{} 
\newcommand{\mab}{\mbox{$m_{AB}$}}

\newcommand{\degsq}{\mbox{deg$^{2}$}}
\newcommand{\degs}{\mbox{$^{\circ}$}}

\newcommand{\si}{\mbox{$\sim$}}
\newcommand{\mm}{\mbox{$\mu$m}}

\newcommand{\ib}{\mbox{\emph{i}-band}}

\newcommand{\kb}{\emph{K}-band}
\newcommand{\kd}{\mbox{\emph{K}$_{d}$}}

\newcommand{\kdb}{\mbox{\emph{K}$_{d}$-band}}

\newcommand{\qb}{\emph{Q}-band}

\newcommand{\ocen}{$\omega$ Cen}

\title{\large\bf\flushleft The Science Case for PILOT I: Summary and Overview}
\author{\parbox{\textwidth}{\flushleft
\vspace{-0.5cm}
{\it J.S.~Lawrence\affil{\,A,AA}, M.C.B.~Ashley\affil{\,A}, J.~Bailey\affil{\,A}, D.~Barrado~y~Navascues\affil{\,B},
T.~Bedding\affil{\,C}, J.~Bland-Hawthorn\affil{\,C}, I.~Bond\affil{\,D}, F.~Boulanger\affil{\,E},
R.~Bouwens\affil{\,F}, H.~Bruntt\affil{\,C}, A.~Bunker\affil{\,G}, D.~Burgarella\affil{\,H}, M.G.~Burton\affil{\,A},
M.~Busso\affil{\,I}, D.~Coward\affil{\,J}, M.-R.~Cioni\affil{\,K}, G.~Durand\affil{\,L},
C.~Eiroa\affil{\,M}, N.~Epchtein\affil{\,N}, N.~Gehrels\affil{\,O}, P.~Gillingham\affil{\,G},
K.~Glazebrook\affil{\,P}, R.~Haynes\affil{\,G}, L.~Kiss\affil{\,C}, P.O.~Lagage\affil{\,L},
T.~Le~Bertre\affil{\,Q}, C.~Mackay\affil{\,R}, J.P.~Maillard\affil{\,S}, A.~McGrath\affil{\,G},
V.~Minier\affil{\,L}, A.~Mora\affil{\,M}, K.~Olsen\affil{\,T}, P.~Persi\affil{\,U},
K.~Pimbblet\affil{\,V}, R.~Quimby\affil{\,W}, W.~Saunders\affil{\,G}, B.~Schmidt\affil{\,X}, D.~Stello\affil{\,C},
J.W.V.~Storey\affil{\,A}, C.~Tinney\affil{\,A}, P.~Tremblin\affil{\,L}, J.C.~Wheeler\affil{\,Y}, and P.~Yock\affil{\,Z}}\\
\vspace{0.4cm}
{\small \affil{A}\,School of Physics, University of New South Wales, NSW 2052, Australia}\\
{\small \affil{B}\,Laboratorio de Astrof\'{\i}sca Espacial y F\'{\i}sica Fundamental (INTA), Madrid 28080, Spain}\\
{\small \affil{C}\,Institute of Astronomy, School of Physics, University of Sydney, NSW 2006, Australia}\\
{\small \affil{D}\,Massey University, Auckland 0745, New Zealand}\\
{\small \affil{E}\,Institut d$'$Astrophysique Spatiale, Universit\'{e} Paris-Sud, Orsay 91405, France}\\
{\small \affil{F}\,Department of Astronomy and Astrophysics, University of California Santa Cruz, Santa Cruz, CA 95064, USA}\\
{\small \affil{G}\,Anglo-Australian Observatory, NSW 1710, Australia}\\
{\small \affil{H}\,Observatoire Astronomique de Marseille Provence, Universit\'{e} d$'$Aix-Marseille, Marseille 13388, France}\\
{\small \affil{I}\,Department of Physics, University of Perugia, Perugia 06123, Italy}\\
{\small \affil{J}\,School of Physics, University of Western Australia, Crawley, WA 6009, Australia}\\
{\small \affil{K}\,Centre for Astrophysics Research, University of Hertfordshire, Hatfield AL10 9AB, UK}\\
{\small \affil{L}\,Service d$'$Astrophysique, CEA Saclay, Saclay 91191, France}\\
{\small \affil{M}\,Departmento de Fisica Te\'{o}rica C-XI, Universidad Aut\'{o}noma de Madrid, Madrid 28049, Spain}\\
{\small \affil{N}\,CNRS-Fizeau/UNSA, Nice 06108, France}\\
{\small \affil{O}\,NASA/Goddard Space Flight Center, Greenbelt, MD 20771, USA}\\
{\small \affil{P}\,Centre for Astrophysics and Supercomputing, Swinburne University of Technology, Hawthorn, VIC 3122, Australia}\\
{\small \affil{Q}\,Observatoire de Paris, Paris 75014, France}\\
{\small \affil{R}\,Institute of Astronomy, University of Cambridge, Cambridge CB3 0HE, UK}\\
{\small \affil{S}\,Institut d$'$Astrophysique de Paris, Paris 75014, France}\\
{\small \affil{T}\,Kitt Peak National Observatory, National Optical Astronomy Observatory, Tucson, Arizona 85719, USA}\\
{\small \affil{U}\,Istituto Astrofisica Spaziale e Fisica Cosmica/INAF, Roma 00100, Italy}\\
{\small \affil{V}\,Department of Physics, University of Queensland, Brisbane, QLD 4072, Australia}\\
{\small \affil{W}\,Astronomy Department, California Institute of Technology, Pasadena, CA 91125, USA}\\
{\small \affil{X}\,Research School of Astronomy and Astrophysics, Australian National University, ACT 2611, Australia}\\
{\small \affil{Y}\,Department of Astronomy, University of Texas, Austin, TX 78712, USA}\\
{\small \affil{Z}\,University of Auckland, Auckland 1142, New Zealand}\\
{\small \affil{AA}\,JSL now at Department of Physics and Electronic Engineering, Macquarie University, NSW 2109, Australia; and Anglo-Australian Observatory, NSW 1710, Australia; Email: jsl@physics.mq.edu.au}}}
\begin{document}
\twocolumn[
\begin{changemargin}{.8cm}{.5cm}
\begin{minipage}{.9\textwidth}
\vspace{-1cm}
\maketitle
%
%
\medskip
\medskip
\end{minipage}
\end{changemargin}
] \small
\twocolumn[ \small{\bf Abstract:} PILOT (the Pathfinder for an International Large
Optical Telescope) is a proposed 2.5~m optical/infrared telescope to be located at Dome~C
on the Antarctic plateau. Conditions at Dome~C are known to be exceptional for astronomy.
The seeing (above $\sim$30~m height), coherence time, and isoplanatic angle are all twice
as good as at typical mid-latitude sites, while the water-vapour column, and the
atmosphere and telescope thermal emission are all an order of magnitude better. These
conditions enable a unique scientific capability for PILOT, which is addressed in this
series of papers. The current paper presents an overview of the optical and
instrumentation suite for PILOT and its expected performance, a summary of the key
science goals and observational approach for the facility, a discussion of the synergies
between the science goals for PILOT and other telescopes, and a discussion of the future
of Antarctic astronomy. Paper II and Paper III present details of the science projects
divided, respectively, between the distant Universe (i.e., studies of first light, and
the assembly and evolution of structure) and the nearby Universe (i.e., studies of Local
Group galaxies, the Milky Way, and the Solar System).

]

\section{Introduction}
The very low infrared background and atmospheric water-vapour content of Antarctic
plateau sites such as Dome~C enables a telescope based there to achieve the sensitivity
(at some wavelengths) of a telescope over three times the diameter located elsewhere
\citep{Lawrence_04,Walden_e_05,Tomasi_e_06}. The atmospheric turbulence above Dome~C is
also 2 to 3 times lower than that at even the best temperate sites
\citep{Lawrence_e_04,Agabi_e_06,Trinquet_e_08}. An optical/infrared telescope at Dome~C
would thus be supremely powerful for its size, enjoying not only a substantial advantage
in both sensitivity and photometric precision, but also having a wide-field,
high-resolution, high-cadence imaging capability otherwise achievable only from space
\citep{Kenyon_Storey_06,Kenyon_e_06,Mosser_Aristidi_07}.

PILOT (Pathfinder for an International Large Optical Telescope) is a key step to a major
international observatory at Dome~C. It is proposed as a high spatial resolution
wide-field telescope with an optical design that is matched to the atmospheric conditions
and a suite of instruments operating at wavelengths from the visible to the mid-infrared
and beyond \citep{Saunders_e_08a,Saunders_e_08b}. The specific aims for PILOT are:
\begin{itemize}
\item to perform cutting-edge science;
\item  to validate and further characterise the expected excellent natural seeing and
    the low thermal sky backgrounds at Dome~C, and to demonstrate that we can fully
    utilise these site conditions;
\item to demonstrate that large optical/infrared telescopes can be built and operated
    in Antarctica within a reasonable time and cost.
\end{itemize}

The PILOT project, and its scientific motivation, has evolved over a number of years.
Early discussions of the scientific potential of Antarctic plateau sites for generic
optical and infrared telescopes were given by \citet{Burton_e_94} and
\citet{Burton_e_01}. A more detailed investigation giving a broad range of potential
science programs for a 2~m class Antarctic telescope was presented by
\citet{Burton_e_05}, based on a ``strawman" telescope and instrument suite configuration
for PILOT. In the current series of papers, we present the next iteration of the PILOT
science case. A series of leading-edge science drivers for the facility have been
identified, and the specific observational and technological requirements have been
defined for each. This science case has evolved in parallel with the telescope optical
design and instrument suite configuration \citep[see][]{Saunders_e_08a,Saunders_e_08b}
developed during the ``Phase A" PILOT design study\footnote{See
\url{http://www.phys.unsw.edu.au/pilot/}}. It is the intention that the science cases and
observing strategies described here, and the requirements that these make on the
telescope and instrument design, will be further refined, iterated, and prioritised
during the next phase of the PILOT project.

The key science objectives that have been identified for the PILOT project, grouped into
seven themes, are:
\begin{enumerate}
\item First light in the Universe: to detect pair-instability supernovae and
    gamma-ray burst afterglows at high redshift. These objects represent the
    signatures of the final evolutionary stages of the first stars to form in the
    Universe.
\item The assembly of structure: to examine the properties of the first evolved
    galaxies to form in the Universe at high redshift, and to examine the assembly
    processes of galaxy clusters at moderate redshift.
\item Dark matter and dark energy: to probe the evolution of cosmological parameters
    via the observation of weak gravitationally-lensed galaxies and supernovae
    unaffected by dust extinction.
\item Stellar properties and populations: to increase our understanding of the
    formation and evolution of galaxies and stars, through investigations of the
    properties of stellar populations in nearby galaxies and stellar clusters.
\item Star and planet formation: to investigate the molecular phase of the Galaxy and     explore the ecology of star formation, and to investigate the formation processes     of stellar and planetary systems.
\item Exoplanet science: to directly detect ``free floating" objects as low in mass
    as a few Jupiter masses, to firmly determine the abundance of ice-giant planets,
    and to characterise the atmospheric properties of a large number of hot Jupiters.
\item Solar system and space science: to investigate the atmospheric composition and dynamics of Venus and the atmospheric surface pressure of Mars, to observe coronal mass ejections from the Sun, and to obtain orbits for a large number of small-scale space debris (also known as ``space junk").
\end{enumerate}

The depth and breadth of science proposed for the PILOT facility, despite its modest
2.5~m aperture, is indicative of the potential of the Dome~C site. These same
characteristics will also benefit future, larger telescopes, for which PILOT will act as
a pathfinder. For example, an 8--10~m class Antarctic telescope would be extraordinarily
powerful for very high resolution optical imaging over small fields and/or deep infrared
imaging over extremely wide fields, enabling direct imaging of exoplanets and detailed
investigations of the first stars and galaxies to form in the Universe.

This paper is organised as follows: Section 2 gives an overview of the Dome~C site
conditions, the PILOT optical design and instrumentation suite, and its expected
performance. An analysis of the PILOT parameter space compared to current and future
facilities is also given. Section 3 summarises, and provides the context for, the key
PILOT science projects. A discussion of observing strategies for the facility is given in Section 4. Synergies between PILOT and other facilities, in particular the Giant Magellan Telescope, the Australian Square Kilometre Array Pathfinder, the Murchison Widefield
Array, and the South Pole Telescope, are addressed is Section 5. Finally, Section 6
presents a vision for the future of Antarctic astronomy via a discussion of potential larger-scale astronomical facilities to follow PILOT. Paper II \citep{Lawrence_e_09a} in this series presents a range of science projects for the PILOT facility that are aimed at observing and understanding the distant Universe (i.e., science themes 1--3 from the above list). Paper III \citep{Lawrence_e_09b} discusses PILOT science projects dealing with the nearby Universe (i.e., science themes 4--7).

\begin{figure}[h]
\begin{center}
\includegraphics[width=7.5cm]{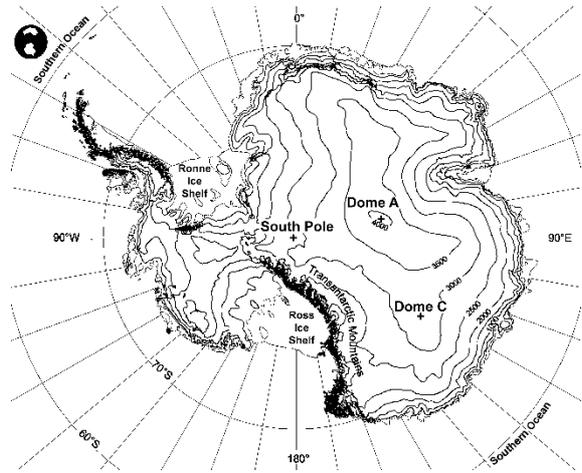}
\caption{Map of Antarctica showing the high plateau stations South Pole, Dome~C and
Dome~A. Basic map courtesy of the Australian Antarctic Data Centre.}\label{map}
\end{center}
\end{figure}

\section{PILOT Design and Performance}

\subsection{Dome C Site Conditions}

Nearly two decades ago it was recognised that the high, dry, cold sites on the Antarctic
plateau should be ideal for astronomy (see \citet{Storey_05} for a recent review).
\citet{Harper_89} suggested that the low temperatures throughout the Antarctic plateau
troposphere would result in a very low atmospheric thermal emission in the infrared.
\citet{Townes_Melnick_06} suggested that the low atmospheric water vapour content would
be particularly suitable for infrared and millimetre wave observations.
\citet{Gillingham_91} predicted that the calm and stable atmosphere above the Antarctic
plateau would result in weak turbulence, and that the strong temperature inversion and
low winds on the domes of the high plateau would result in the atmospheric turbulence
being confined to a very thin but intense layer close to the surface.

Since that time, site-testing experiments have been carried out at, or are planned for, a number of locations (South Pole, Dome~C, Dome~A, and Dome~F) on the Antarctic plateau
(see Figure~\ref{map}). These results have confirmed and quantified the earlier
expectations, and have demonstrated that each of these sites has distinct
characteristics, and is therefore most appropriate to a specific astronomical field.
Dome~C, the site of the French/Italian Concordia station \citep[see][]{Candidi_Lori_03},
has been selected as the most appropriate location on the plateau for the PILOT
telescope.

Dome~C, at 123\degs\ East longitude and 75\degs\ South latitude, is at an altitude of
3250~m. It has been operated by the French and Italian Antarctic agencies, Institut Paul
Emile Victor (IPEV) and Programma Nazionale di Ricerche in Antartide (PNRA), during the
summer months since 1996, and has been manned year round since 2005.

The key atmospheric parameters and site conditions at Dome~C relevant to the scientific
potential for PILOT are\footnote{The \emph{PILOT Dome~C Environmental Conditions
Document} (and references therein) describes in detail the meteorological atmospheric
conditions and astronomical site conditions at Dome~C. See
\url{http://www.phys.unsw.edu.au/pilot/pilot_status.htm}.}:
\begin{itemize}
\item excellent free-atmospheric seeing
    \citep{Lawrence_e_04,Agabi_e_06,Trinquet_e_08},
\item low turbulent boundary layer height
    \citep{Lawrence_e_04,Agabi_e_06,Trinquet_e_08},
\item wide isoplanatic angle and long coherence time
    \citep{Lawrence_e_04,Agabi_e_06,Trinquet_e_08},
\item low atmospheric scintillation \citep{Kenyon_e_06},
\item low sky and telescope thermal emission \citep{Lawrence_04,Walden_e_05},
\item low precipitable water vapour content
    \citep{Valenziano_DallOglio_99,Lawrence_04,Tomasi_e_06},
\item high cloud-free fraction \citep{Mosser_Aristidi_07}, and
\item high latitude \citep{Kenyon_Storey_06}.
\end{itemize}

\subsection{PILOT Configuration}
\begin{figure}[h]
\begin{center}
\includegraphics[width=7.5cm]{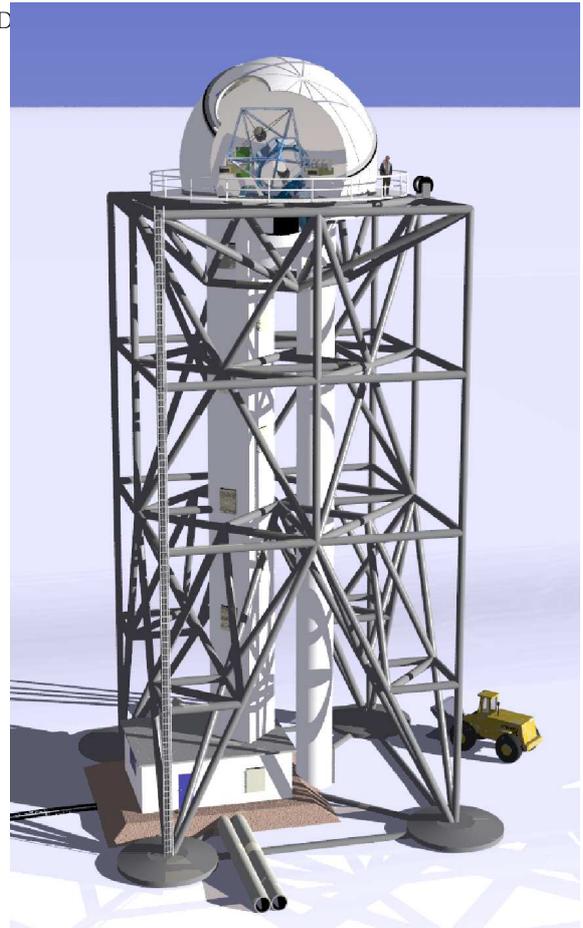}
\caption{Concept design for the PILOT telescope, enclosure, and tower.}\label{pilot_tower}
\end{center}
\end{figure}

The baseline optical design for PILOT, described in \citet{Saunders_e_08a}\footnote{See
also the \emph{PILOT Functional and Performance Requirements Document} at
\url{http://www.phys.unsw.edu.au/pilot/pilot_status.htm.}}, comprises a 2.5 m
Ritchey-Chretien telescope with f/1.5 primary and f/10 overall focal ratios, giving
diffraction-limited performance at 1~\mm\ over a 1\degs\ field. Instruments are mounted
on twin Nasmyth foci. The telescope is housed in a calotte-style dome that is mounted on
top of a $\sim$30~m high tower in order to elevate the main mirror above the majority of
the intense ground-layer turbulence (see Figure~\ref{pilot_tower}). The enclosure is
temperature and humidity controlled, protecting the optical elements from large spatial
and temporal thermal gradients, and preventing frost formation on optical surfaces
\citep{Durand_e_07}. A fast tip-tilt secondary mirror is used for guiding and to remove
residual boundary-layer turbulence and tower wind-shake. The system is designed for
continuous 24~hour remote operation with minimal human intervention.

\subsection{PILOT Baseline Instruments}
\begin{table*}[t]
\begin{center}
\caption{PILOT baseline instruments, their main parameters, and primary science
drivers.}\label{Table1}
  \begin{tabular} [t] {l l l l l}
\hline
Instrument      &	Primary science	  & Pix scale & $\lambda$ range 	 & FOV\\
                &                                         & ($''$/px) &  (\mm)   \\
\hline
PVISC: Visible Camera        &   weak lensing &   0.08      & 0.4--1    & 40$'\times40'$\\
& (paper II) \\
\hline
PNIRC: Near-Infrared           & first light, high z gal   &  0.06     & 1--2.5    & 4$'\times4'$  \\
Camera   &  (paper II)  &  0.15     & 1--5      & 10$'\times10'$ \\	
\hline
PMIRIS: Mid-Infrared          & Galactic ecology     &  0.8     & 7--25    & 14$'\times14'$ \\
Imaging Spectrometer    &    (paper III)              & 1.3  & 17--40   & 5$'\times5'$ \\
\hline
PLIC: Lucky Imaging           & ``Hubble from the       & 0.03    & 0.4--1    & 0.5$'\times0.5'$  \\
Camera   & ground" (paper I)         \\

\hline
\end{tabular}
\medskip\\
\end{center}
\end{table*}

The following list shows the types of astronomical observations where PILOT can
outperform a similar or somewhat larger telescope at a temperate latitude observatory.

\begin{itemize}
\item High-resolution near-diffraction-limited imaging in the visible over small
    fields (resulting from the excellent free-atmospheric seeing, the wide
    atmospheric isoplanatic angle, and the long atmospheric coherence time).
\item Moderate-resolution wide-field imaging in the visible and near-infrared with
    partial (tip-tilt) correction of the residual boundary-layer turbulence
    (resulting from the excellent free-atmospheric seeing, the low height of the
    turbulent boundary layer, the wide atmospheric isoplanatic angle, and the long
    atmospheric coherence time), and moderate-resolution wide-field imaging in the
    mid-infrared (limited by the telescope aperture rather than the atmospheric
    conditions).
\item High sensitivity in the near-infrared (arising from the low atmospheric
    thermal emission) and the mid-infrared (arising from a combination of the low
    atmospheric thermal emission and the high atmospheric transmission).
\item High photometric precision in the optical (enabled by the low atmospheric
    scintillation) and the infrared (enabled by the stable atmospheric thermal emission).
\item Continuous coverage (due to the high latitude of the Dome~C site and the high
    cloud-free fraction).
\end{itemize}

The baseline instrument suite for PILOT has been designed to take advantage of all the
above opportunities. The suite consists of four cameras, of which three are imaging only
and one has spectro-photometric capabilities. All four instruments will be permanently
mounted on the telescope Nasmyth platform in such a way that they are individually
selectable via a tertiary fold-mirror.

The four instruments, described in detail by \citet{Saunders_e_08b}, are:
\begin{itemize}
\item PVISC (PILOT VISible Camera): a wide-field optical camera with ground-layer
    tip-tilt correction.
\item PNIRC (PILOT Near-InfraRed Camera): a wide-field near-infrared camera with
    ground-layer tip-tilt correction and adjustable pixel scales matched to the
    diffraction limit at short and long wavelengths.
\item PMIRIS (PILOT Mid-InfraRed Imaging Spectrometer): a wide-field mid-infrared
    instrument with a tuneable Fabry-Perot filter or a GRISM spectrometer, and two
    separate arms with short and long wavelength ranges.
\item PLIC (PILOT Lucky Imaging Camera): a fast optical camera for
    diffraction-limited imaging over relatively small fields in the visible.
\end{itemize}

The key parameters for each of these instruments, and the appropriate primary science drivers, are given in Table ~\ref{Table1}. The specifications given
here should be considered indicative; a range of applicable values for each instrument
specification has been identified\footnote{See the \emph{PILOT Science Requirements
Document} and the \emph{PILOT Functional and Performance Requirements Document} at
\url{http://www.phys.unsw.edu.au/pilot/pilot_status.htm.}} and awaits a more detailed
trade-off analysis before selection of the optimum configuration.

\subsection{Advanced Technology Instruments}
In addition to the baseline instruments defined in the previous section, there are
several ``advanced technology" instruments that are also being considered. These are
either instruments for a generic \si2~m class Antarctic telescope that are being
developed independent to the PILOT project, or instrument configurations that involve a
significant extension to the baseline specifications. They include:

\begin{itemize}
\item SmilePILOT: a sub-millimetre imaging camera. The SmilePILOT instrument consists     of a filled-bolometer array camera operating from 200--450~\mm\ \citep{Minier_e_07,Minier_e_08}. At 200~\mm, it will provide a FWHM beam of 21$''$, with a total field-of-view of $9.2'\times9.2'$.
\item AIFU (Antarctic Integral Field Unit): a near-infrared integral field
    spectrograph fed with a fibre array \citep{Le_Bertre_e_08}. This instrument samples a $1'\times1'$ field-of-view at 2$''$, giving a spectral resolution of $R\approx100$ covering the wavelength range 1--2.5~\mm.
\item Polar Bear: an Imaging Fourier Transform Spectrometer (IFTS) to operate in the
    near-infrared \citep{Maillard_Boulanger_08}. Polar Bear consists of a dual-arm
    interferometer sampling a 10$'\times10'$ field-of-view at 0.3$''$ operating over the wavelength range 1.8--5.5~\mm, with a maximum spectral resolution of $R=125,000$.
\item PSDC (PILOT Satellite Debris Camera): a wide-field imaging camera for satellite     debris tracking. This instrument, which consists of a large mosaic of fast-readout CCDs, should obtain high spatial resolution imaging in the visible over a wide field (1~\degsq).
\end{itemize}

\subsection{PILOT Performance}
The design for the PILOT telescope and instrument suite must ultimately be driven by the
requirements determined by the scientific objectives. These objectives can only be
defined based on some initial specification. The starting point for the science case is
the baseline instrument suite, described in Section 2.3, and its expected performance,
described here.

The estimated system performance, i.e., achieved resolution and sensitivity, is given in
Table ~\ref{Table2}, based on the analysis of \citet{Saunders_e_08a}. The image quality
specification includes telescope aberrations and surface errors; misalignment errors;
telescope, mirror, and dome seeing; residual tower wind-shake; and guiding errors
introduced via the tip-tilt measurement and correction system. The sensitivity analysis
includes sky and telescope thermal emission, and assumes optimum source extraction. The
expected resolution and sensitivity for PILOT depends on many site parameters
(particularly the free-atmospheric seeing, the turbulence profile, and the thermal
infrared background) that are not completely characterised, and also on details of the
telescope design (e.g., plate scale, baffling, guiding system) that are not yet
finalised. The estimated system performance, as summarised in Table~\ref{Table2}, is thus
intended to be indicative rather than definitive.

\begin{table*}[t]
\begin{center}
\caption{The expected resolution and sensitivity for PILOT in a number of
wavebands.}\label{Table2}
\begin{tabular}{llccccccc}
\\
\hline Band & $\lambda$ & $R$ & FWHM $^a$ & \mab $^b$ & $m_{Vega}$ $^b$ &\mab $^b$ & $m_{Vega}$ $^b$\\
 & (\mm) & ($\lambda/\Delta\lambda)$ & (arcsec) &  & & arcsec$^{-2}$& arcsec$^{-2}$\\
\hline
$g$                 &   0.47      & 3.4 &	0.35 &	27.6 & 27.6	& 27.1 & 27.1\\
$r$                 &	0.62      & 4.4 &	0.33 &	27.1 & 26.9	& 26.5 & 26.3\\
$i$                 &	0.76      & 5.1 &	0.32 &	26.6 & 26.2	& 26.0 & 25.6\\
$z$                 &	0.91      & 6.5 &	0.31 &	25.8 & 25.3	& 25.1 & 24.6\\
$Y$                 &	1.04      & 5.1 &	0.30 &	25.5 & 24.9	& 24.8 & 24.2\\
$J$                 &	1.21      & 4.6 &	0.30 &	25.0 & 24.1	& 24.3 & 23.4\\
$H$                 &	1.65      & 5.7 &	0.29 &	24.6 & 23.2	& 23.8 & 22.4\\
$K_{d}$             &	2.40$^c$  & 10  &	0.32 &	25.3 & 23.3 & 24.7 & 22.7\\
$L$                 &   3.76      & 5.8 &	0.40 &	21.2 & 18.3	& 20.8 & 17.9\\
$M$                 &   4.66      & 19  &	0.46 &	19.6 & 16.2	& 19.4 & 16.0\\
$N^\prime$           &   11.5      & 11  &	1.05 &	16.3 & 11.2	& 17.0 & 11.9\\
$Q_{N}$             &   20.1      & 20  &	1.80 &	14.6 & 8.1	& 15.8 & 9.3\\
\hline
\end{tabular}
\medskip\\
\end{center}
$^a$ The resolution full-width-half-maximum (FWHM) over the full imaging field-of-view
for each camera, is given as a function of wavelength based on a preliminary analysis of
the efficiency of the proposed ground-layer tip-tilt correction system.
For wavelengths longer than \si3~\mm\ the corrected resolution is close to diffraction limited.\\
$^b$ Point-source and extended-object limiting sensitivities (in magnitudes and
magnitudes per square arcsec respectively) are given for a 5~$\sigma$, 1~hour
integration, assuming that the sky background is summed over 4 times the FHWM disc (for
point sources), the telescope is at 227~K with 5\% emissivity, the overall optical
efficiency is 50\% (including throughput, detector efficiencies, and secondary mirror
obscuration) and the infrared sky backgrounds are as
given in \citet{Burton_e_05}.\\
$^c$ The \kdb\ (or $K_{dark}$-band) is defined with a centre wavelength of 2.4~\mm. This band is centred at a slightly higher wavelength than the standard \kb, in order to take maximum advantage of the low thermal emission of the Antarctic atmosphere. \\ \\
\end{table*}

\subsection{PILOT Discovery Space}

\begin{figure*}[h!]
\begin{center}
\includegraphics[width=14cm]{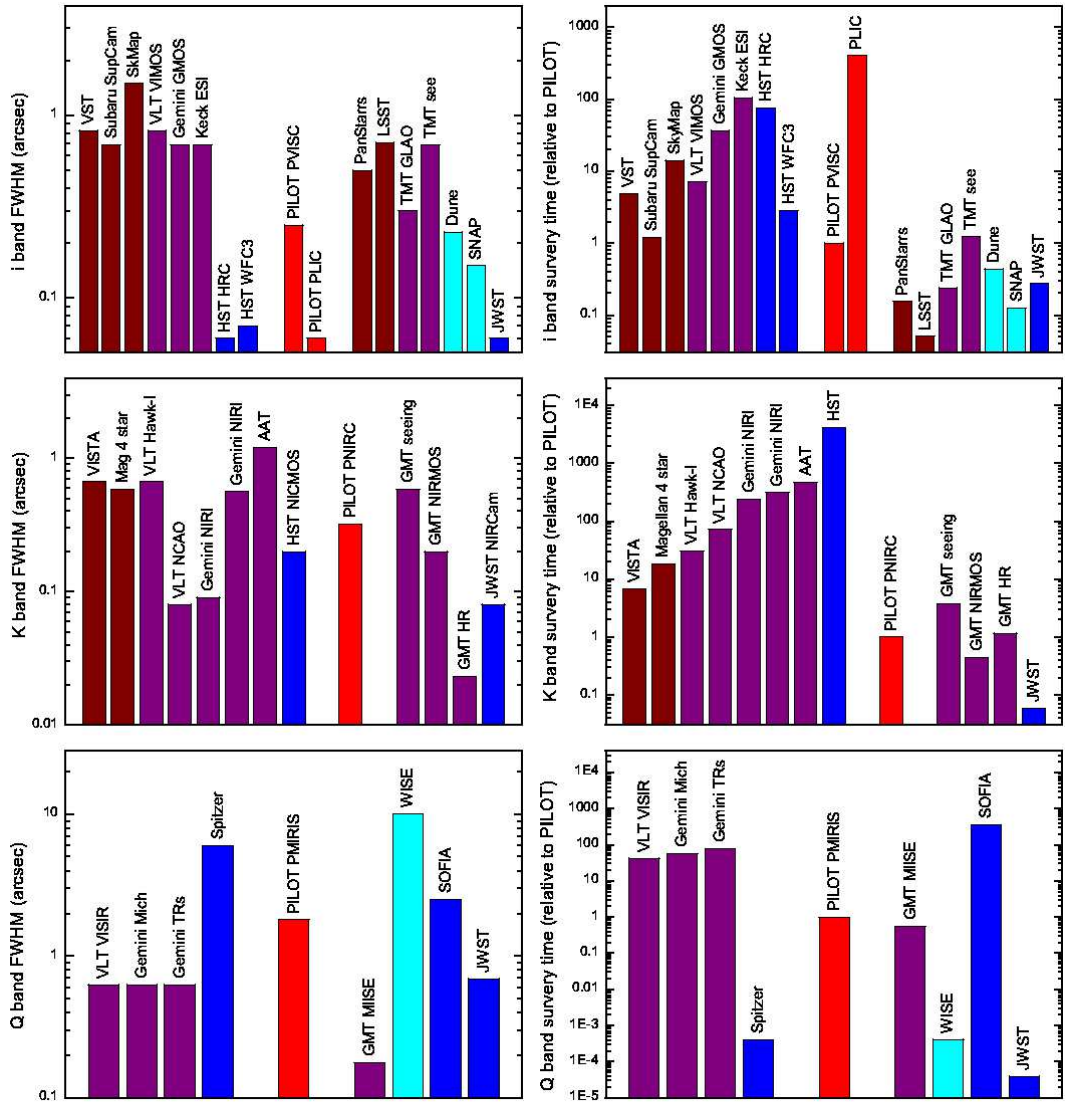}
\caption{Resolution (left column) and survey time relative to PILOT (right column) for a
range of facilities at \ib\ (top row), \kb\ (middle row), and \qb\ (bottom row). Survey
time represents the time required to survey a given area of sky to a given depth. For
both FWHM and survey time a lower number is preferable (i.e., giving a high spatial
resolution in a short amount of observing time). The comparison includes wide-field
ground-based survey telescopes or instruments (brown), ground-based general-purpose
telescopes (purple), space-based wide-field survey telescopes (light blue), and
spaced-based general-purpose telescopes (dark blue). Existing and near-complete facilities are shown to the left of PILOT, and future planned or proposed facilities to the right of PILOT.}\label{discovery}
\end{center}
\end{figure*}

The scientific capabilities for PILOT need to be considered in the context of the
capabilities of other current, planned, or proposed facilities. Figure~\ref{discovery}
shows the resolution and relative survey time (ratio of integration times required to
image a given area of the sky to a given sensitivity) for PILOT compared to a number of
other facilities in the visible (\ib), near-infrared (\kb), and mid-infrared (\qb). One of the key benefits of PILOT is that it will operate over this entire wavelength range. The relative efficiency metric used here includes, for each telescope, pixel scale, resolution (either seeing-limited or corrected using adaptive optics), telescope effective diameter, atmospheric (or space) thermal emission, telescope thermal emission, and instrument field-of-view.\footnote{Telescope, instrument, and site parameters taken from
\citet{Kaiser_e_02,Fazio_e_04,Rhodes_e_04,Rieke_e_04,Sholl_e_04,Dalton_e_06,Mainzer_e_06,Tokovinin_Travouillon_06,Werner_e_06,Keller_e_07,Thomas_Osip_07,Ivezic_e_08},
and \url{http://www.gemini.edu/sciops/instruments/};
\url{http://www.gmto.org/CoDRpublic};
\url{http://www.eso.org/sci/facilities/paranal/instruments};
\url{http://www.eso.org/gen-fac/pubs/astclim/paranal/seeing/};
\url{http://wise.ssl.berkeley.edu/mission.html}; \url{http://www.spitzer.caltech.edu/};
\url{http://casu.ast.cam.ac.uk/documents/vista/}; \url{http://www.tmt.org/index.html};
\url{http://vstportal.oacn.inaf.it/node/1}; \url{http://www.stsci.edu/hst/};
\url{http://www.dune-mission.net/}; \url{http://snap.lbl.gov/};
\url{http://www.lsst.org/lsst_home.shtml};
\url{http://pan-starrs.ifa.hawaii.edu/public/home.html}} In Figure~\ref{discovery} we show both general purpose and survey telescopes.  As expected, PILOT is usually (though not always) faster in a survey mode than telescopes that are not designed with surveys in mind.

In the visible, PILOT provides a higher resolution (not including Lucky Imaging systems)
and a faster survey speed than any current ground-based telescope. HST (Hubble Space
Telescope) provides a higher resolution, but is slower. Future wide-field optical
facilities such as Pan-STARRS-4 (Panoramic Survey Telescope And Rapid Response System)
and LSST (Large Synoptic Survey Telescope) are a factor 10--40 times faster than PILOT
but will not achieve the same spatial resolution. PILOT is faster at imaging and has a
higher resolution than ground based ELTs (Extremely Large Telescopes) in seeing-limited
mode, but loses the advantage if GLAO (Ground Layer Adaptive Optic) systems can be
routinely operated on those facilities. It is unlikely, however, that ELTs will do large
imaging surveys, as they are better suited to narrow spectral resolution work in this
wavelength range. The proposed Euclid, previously DUNE (Dark UNiverse Explorer), and SNAP (SuperNova/Acceleration Probe) space missions provide slightly better spatial resolution
and are 4--12 times faster than PILOT. They are not due for first light until at least 5
years after PILOT; it is thus possible that PILOT could achieve many of their proposed
science goals before they are launched. The planned visible camera for JWST (James Webb
Space Telescope) provides a factor of \si6 in imaging speed and has a much higher spatial resolution than PILOT. However, JWST is optimised for performance in the thermal infrared and it is currently not expected that large projects in the visible will be undertaken.

The PILOT niche in the near-infrared at \kb\ arises from a combination of the relatively
high spatial resolution and the low thermal background. PILOT is at least 8 times faster to image a given area of sky than any current telescope at this wavelength, and only HST and adaptive-optics equipped 8--10~m class telescopes can compete for
resolution. PILOT is either faster with a higher resolution, or slower with a lower
resolution than the GMT (Giant Magellan Telescope), depending on the GMT instrument
considered (primarily related to the adaptive optics capabilities). As with the optical,
however, the key ELT role in the near-infrared will be high spectral and/or spatial
resolution observations. JWST will the pre-eminent telescope in the near-infrared. It is
diffraction limited at these wavelengths. As shown here JWST is 10--500 times faster than
any existing ground based telescope, and is \si20 times faster than PILOT at \kb.
However, there is a significant efficiency loss associated with JWST telescope slewing
and settling, which precludes its use for large-area surveys. Its real niche lies in
obtaining deep high-resolution images over relatively small regions of sky.

The niche for PILOT in the mid-infrared wavelength range is clear. It falls in the large
gaps between ground and space telescopes in terms of survey speed, and between
large-aperture small-field telescopes and small-aperture wide-field telescopes in terms
of resolution. It is the only telescope capable of obtaining moderate spatial resolution
photometry over very wide regions of sky, and is thus an ideal complement to both the
deep narrow fields of JWST and GMT and the wide-area low-spatial-resolution fields of
Spitzer and WISE (Wide-field Infrared Survey Explorer).

\subsection{Lucky Imaging with PILOT}
The PILOT project has considerable potential to extend the parameter space demonstrated
by Lucky Imaging camera systems. The unique site characteristics at Dome~C should allow
the proposed PILOT Lucky Imaging Camera (PLIC) to achieve higher resolution over a larger fraction of the sky at shorter wavelengths than other Lucky Imaging systems operated on
mid-latitude telescopes. PLIC is expected to play an important ``pathfinding" role for
the PILOT project, where the delivered performance will provide a confirmation of the
atmospheric conditions of the site; there is no better way to verify these atmospheric conditions. The technique is therefore described in more detail
here.

\begin{figure}[t]
\begin{center}
\includegraphics[width=2.5cm]{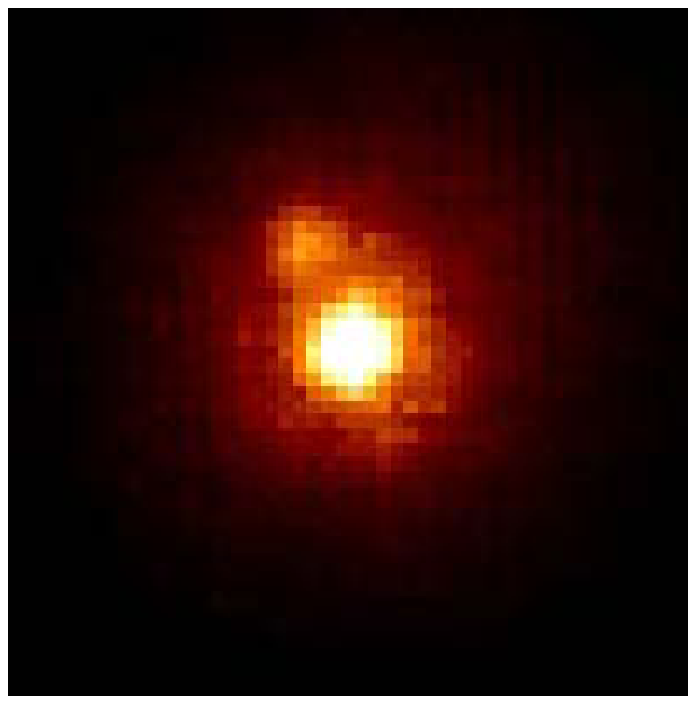}
\includegraphics[width=5.0cm]{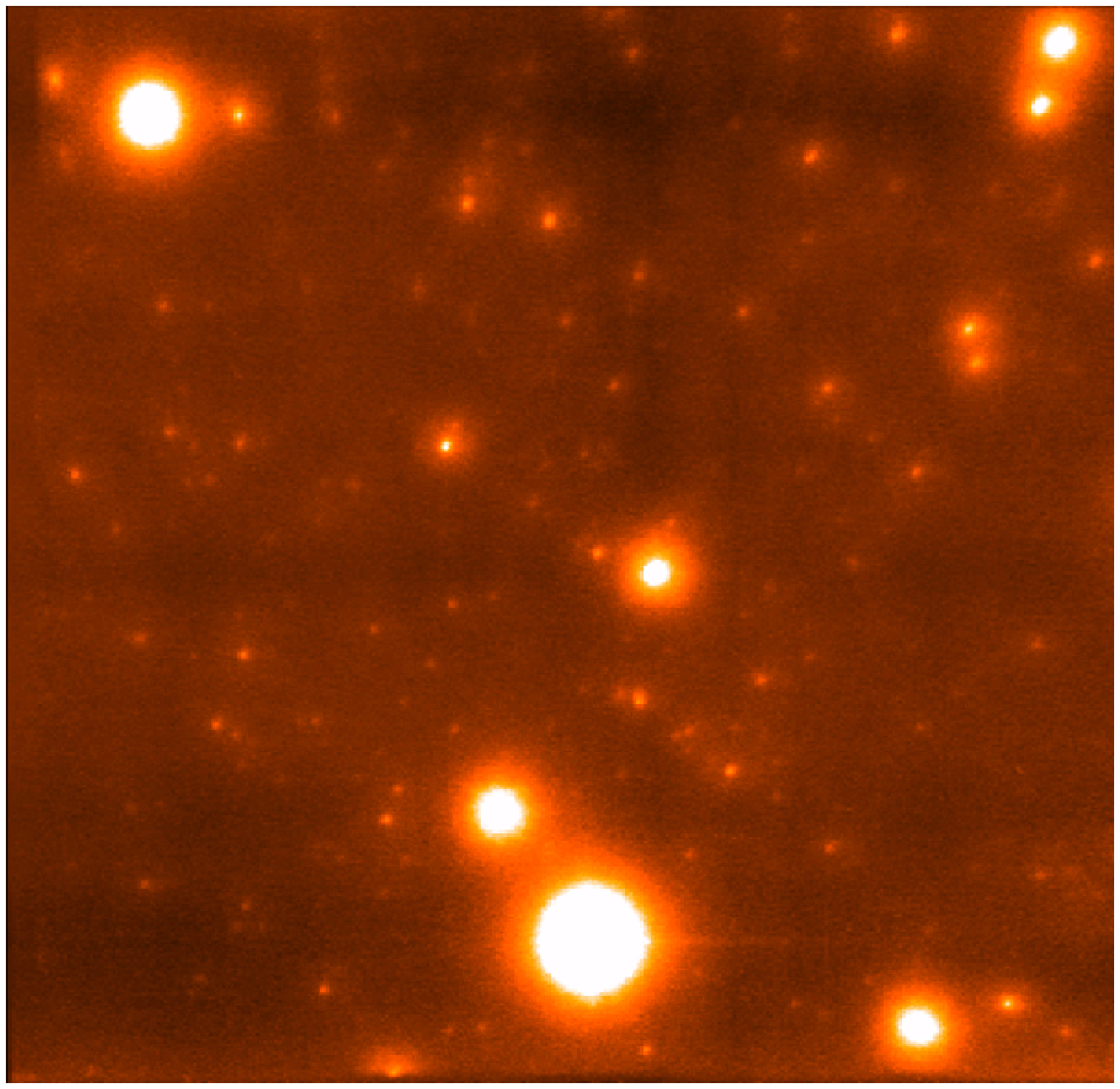}
\caption{Left: 0.12$''$ binary, $\delta\mathrm{mag}\approx2.5$, imaged with LuckyCam on the NOT
2.5~m telescope. Right: the highest resolution optical image ever taken, with a
resolution of 0.035$''$ in \ib, using the Lucky Camera behind the Palomar 5~m telescope
low-order adaptive optics system. From \citet{Mackay_e_08}. A similar resolution should
be achievable with PLIC on PILOT at wavelengths around 0.5~\mm. This is \si2 times higher
resolution than possible with HST.}\label{LI}
\end{center}
\end{figure}

The Lucky Imaging technique, first proposed by \citet{Fried_78}, aims to improve the
angular resolution of astronomical instruments on ground-based telescopes. The process
involves the collection of sequences of short-exposure frames. The frames are then sorted
by Strehl ratio. The selected fraction are then re-centred on the brightest speckle and
added. Non-selected frames may be used to increase sensitivity at the expense of
resolution. Lucky Imaging has been demonstrated to be routinely usable on 2.5~m class
optical telescopes on good high-altitude sites and able to produce spatial resolutions of
\si0.1$''$ in \ib\ \citep{Law_e_08} using reference stars as faint as $\mab = 16.3$--16.8
\citep{Law_e_06}. In the near-infrared, deep depletion CCDs will soon be available which
should offer an improvement in sensitivity of about one magnitude. With a suitable
reference star, the faint limit for science targets is that predicted for the selected
image integration time.

As the Lucky Imaging technique involves the statistics of atmospheric turbulence, the
expected performance of such systems is a strong function of the atmospheric site
conditions \citep{Baldwin_e_08}. The good seeing and long coherence time of the Dome~C
atmosphere should allow the system to work with a slower frame rate, improving the
reference star sensitivity. A good working estimate for the reference star limiting
magnitude for PLIC is probably $\mab\approx18.3$ at \ib, corresponding to a star density
that gives a 50\% chance of finding a reference star within a 1~arcmin$^{2}$ field at
high galactic latitudes. Lucky Imaging also benefits from a large isoplanatic patch
\citep{Baldwin_e_08}. The isoplanatic angle is typically 1$'$ at \ib\ on a site such as
La Palma; the expectation is that it should be a factor 2--3 larger at Dome~C. There is
thus a higher probability of finding a suitable reference star within the isoplanatic
patch of PILOT. In addition, ground-layer turbulence effects are negated because each
frame is shifted and added before recombination. There is a significant sensitivity hit
for Lucky Imaging, as the ``bad" frames are thrown away. To obtain reasonable Strehl
ratios from mid-latitude sites typically requires selection of the best 1--10\% of images. The number of retained images is a strong function of the atmospheric seeing, and thus, for PLIC we may expect that as many as 50\% of frames can be retained,
corresponding to a significant gain in sensitivity.

All of the performance metrics for Lucky Imaging (sky coverage factor, reference star
limiting magnitude, number of images collected, and achieved resolution) degrade with a
decrease in wavelength. PLIC should be capable of obtaining better performance than a
mid-latitude Lucky Imaging system operating at the same wavelength. Alternatively, PLIC
should achieve similar performance at shorter wavelengths, see Figure~\ref{LI}. PLIC
should, in theory, be capable of delivering a higher resolution than the HST actually
delivers in the \emph{B} and \emph{V} bands\footnote{See the Wide Field Camera 3
Instrument Handbook at \url{http://www.stsci.edu/hst/.}}.

\section{The PILOT Science Case}

\subsection{Context}
Four or five years from now, when PILOT could be operational, there will have undoubtedly
been many developments in astronomy. Here we summarise the potential astronomical
facility landscape at that time, assuming that currently planned projects are completed
on schedule.

On the ground, a new generation of instruments will be available on current 8--10 m class
telescopes and multi-conjugate adaptive optics systems using multiple laser guide stars
will be operating efficiently and reliably. Ground Layer Adaptive Optics (GLAO) systems
on smaller-scale telescopes now being developed will have obtained useful results. The
VISTA (Visible \& Infrared Survey Telescope for Astronomy), VST (VLT Survey Telescope),
and Skymapper survey telescopes, after operating for several years, will have mapped
large regions of the sky at optical and infrared wavelengths. Currently proposed large
optical survey telescopes, LSST and Pan-STARRS-4, will be nearing completion, and one or
more of the current ELT projects will be fully funded and well into their construction
phase.

At longer wavelengths, the ALMA (Atacama Large Millimeter Array) sub-millimetre array
will have achieved first light, a site will have been selected for the Square Kilometre
Array (SKA) radio array, and SKA prototype facilities, such as the Australian Square
Kilometre Array Pathfinder (ASKAP) and the Murchison Widefield Array (MWA), will be fully operational.

In space, JWST will have achieved first light, NASA will have selected the Joint Dark
Energy Mission (JDEM) project from current mission concepts (SNAP, ADEPT, Destiny), and
ESA will have selected a Cosmic Visions project from current candidates (such as Euclid). The HST may still be operational, and Kepler and Corot will have collected many years of
data on the transient optical sky. WISE and Spitzer will have mapped significant areas of the sky in the mid-infrared.

This array of new facilities will allow new observational possibilities, opening up new
areas of astronomical study. While our understanding of the Universe will thus be
significantly increased, there will no doubt continue to be a vast array of unanswered
questions in all fields of astronomy.

Despite the large scope of these new facilities, the capabilities of PILOT will be
unique; it will provide the highest spatial resolution observations (at any wavelength)
of any wide-field survey-telescope and it will operate at wavelengths from the visible to the mid-infrared. These capabilities will allow PILOT to address some of the fundamental
questions in astronomy. PILOT will allow us to study the structure and composition of the first stars to form in the Universe and the environments in which they form; the history
of reionisation in the Universe and the characteristics of distant galaxies and clusters; the nature and evolution of dark matter and dark energy; the internal properties of stars and the stellar populations of nearby galaxies; the molecular phase of the Galaxy and the processes of star and planet formation; the physical and chemical properties of
exoplanets; and the characteristics of planetary atmospheres.

\subsection{Science Case Summary}
Here we summarise the key science projects for the PILOT facility. More detail is given
in the companion papers in this series. Paper II addresses projects under the themes
\emph{first light in the Universe}, \emph{the assembly of structure}, and \emph{dark
matter and dark energy}. Paper III addresses projects under the themes \emph{stellar
properties and populations}, \emph{star and planet formation}, \emph{exoplanet science},
and \emph{solar system and space science}. In Table~\ref{Table3} we list indicative wavebands, sky coverage, required observing time for the main science projects.

\textbf{First light in the Universe:} a near-infrared search for pair-instability
supernovae (via a dedicated wide-field survey) and gamma-ray burst afterglows (via alerts from high-energy satellites), events which represent the final evolutionary stages of the first stars to form in the Universe.

\textbf{The assembly of structure:} a deep and wide survey in the near-infrared to study
galaxy structure, formation, and evolution via the detection of a large sample of
high-redshift galaxies; and a study of a sample of moderate-redshift galaxy clusters in
the optical aimed at understanding galaxy-cluster growth, structure, and evolution.

\textbf{Dark matter and dark energy:} a wide-area optical survey that will probe the
evolution of dark matter and dark energy via the measurement of the ellipticities of a
large sample of weak gravitationally-lensed galaxies; and a near-infrared search for Type Ia supernovae to obtain light curves that are largely unaffected by dust extinction and
reddening, allowing tighter constraints to be placed on the expansion of the Universe.

\textbf{Stellar properties and populations:} an optical/ near-infrared survey of disc
galaxies in the Local Group to study the processes of galaxy formation and evolution; an
infrared survey of nearby satellite galaxies to trace their outer morphology, structure,
age, and metallicity; a deep mid-infrared survey of the Large and Small Magellanic clouds in order to understand star formation processes; and long time-series optical
observations of nearby globular and open clusters to study age-metallicity relationships, test various predictions of stellar astrophysics, and improve our understanding of the
physics of massive stars.

\textbf{Star and planet formation:} a southern Galactic Plane survey of molecular
hydrogen in the mid-infrared in order to further our understanding of the ecology of star formation; and a series of mid-infrared spectrophotometric surveys searching for
signatures of embedded protostars, crystalline silicates, and circumstellar discs around
young stellar objects and brown dwarfs.

\textbf{Exoplanet science:} a near-infrared search for free-floating planetary mass
objects in nearby stellar clusters; a program for optical follow-up of gravitational
microlensing candidates based on alerts from dedicated survey telescope networks; and a
study of high precision photometric near- and mid-infrared light curves of previously
discovered exoplanets.

\textbf{Solar system and space science:} a series of optical and near-infrared studies aimed at characterising the composition and dynamics of planetary atmospheres of Venus and Mars; a study of the Sun at mid-infrared and sub-millimetre wavelengths aimed at understanding the physical mechanisms that are responsible for coronal mass ejections and solar flares; and a monitoring program aimed at determining orbits for a large number of small Low Earth Orbit satellite debris items.

\begin{table*}[t]
\begin{center}
\caption{Indicative parameters (wavebands, sky coverage, and required observing time) for the main PILOT science projects.}\label{Table3}
\begin{tabular}{llrr}
\\
\hline
theme: project & wavebands & sky area  & observing \\
               &            & (\degsq) & time (hrs)\\
\hline
first light: PISN               & $K_d$         & 40        &	6000 \\
first light: GRB                & $J,H,K_d,L,M$ & ToO       &	200 \\
structure: high z galaxies      & $K_d$         & 3         &	500 \\
structure: clusters             & $r,i,z$       & 40        &	300 \\
dark energy: weak lensing       & $r+i+z$       & 4000      &	2500 \\
dark energy: SNIa               & $J,H,K_d$     & 6         &	2500 \\
stellar pops: Local Group       & $g,r,J,K_d$   & 2         &	1000 \\
stellar pops: satellites        & $Y,J,K_d$     & 5         &	4000 \\
stellar pops: LMC               & $L,K_d$       & 50        &	2000 \\
stellar pops: asteroseismology  & $r$           & 2         &	1000 \\
star formation: H$_2$ survey    & 12,17~\mm     & 10        &	6000 \\
star formation: Chamaeleon      & $N,Q$         & 64        &	3500 \\
exoplanets: free-floaters       & $J,H,K_d$     & 10        &	1000 \\
exoplanets: microlensing        & $g,r$         & ToO       &	750 \\
solar system: planets           & $g,z,K_d$     & 1         &	500 \\
\hline
\end{tabular}
\medskip\\
\end{center}
ToO = Target of Opportunity single-object observations.\\
\end{table*}

\subsection{A PILOT Mission}

The PILOT science case is derived from a diversity of science objectives within a wide
range of fields under the themes just discussed. There are several projects, however,
that we have identified to represent the ``flagship" science for the PILOT facility.
These four scientific projects have a realistic potential to deliver high-impact
breakthrough discoveries, and have been the priority drivers for the telescope optical
and instrumentation suite design:
\begin{enumerate}
\item Pair-instability supernovae and gamma-ray burst afterglow searches (Paper II:
    Sections 2.1 and 2.2)

\item High-redshift galaxy survey (Paper II: Section 3.1)

\item Weak lensing (Paper II: Section 4.1)

\item Galactic ecology (Paper III: Section 3.1).
\end{enumerate}

\subsection{The Pathfinding Role}
A secondary objective for PILOT is to act as a ``path-finder" facility. Firstly, as at all sites, it is not until a telescope large enough to exploit the scientific opportunities has been deployed that it is possible to truly understand the
potential---and limitations---of the site. Secondly, PILOT should demonstrate solutions to the technological and engineering challenges that result from the extreme environmental conditions at Dome~C. These include, for example, the problem of diamond dust and frost accumulation on optical surfaces. Thirdly, PILOT should demonstrate the logistical and operational feasibility of setting up and running an optical/infrared observatory at the Dome~C site, where access is restricted to only a few months per year, and there are limited on-site technical and support staff and facilities.

There are already several larger-scale future Dome~C facilities proposed to follow PILOT
that rely on PILOT having demonstrated the above pathfinder objectives (see Section 6). These include LAPCAT (Large Antarctic Plateau Clear-Aperture Telescope) an 8.4 m diameter off-axis optical/infrared telescope with a primary science driver to directly image exoplanet in the thermal infrared \citep{Storey_e_06}; and IRAST (InfraRed Antarctic Survey Telescope) proposed here as a large aperture, 8--10~m class, wide-field infrared survey telescope.

Future large-scale Antarctic facilities will rely on the successful implementation and
operation of PILOT in much the same way that PILOT will rely on previous Antarctic
infrared telescope projects, such as the 0.6~m South Pole InfraRed Explorer
\citep[SPIREX;][]{Fowler_e_98}, which was operated at the South Pole from 1994--1999.
SPIREX was initially used for site testing, i.e., to quantify the conditions for infrared observations \citep{Nguyen_e_96}. It was then used with a 1--2.5~\mm\ imager
\citep[1994--1997; see ][]{Herald_e_90}, before being equipped with a 2.4--5~\mm\ imager
from 1998--99. As summarised in \citet{Rathborne_Burton_05}, SPIREX undertook a number of science programs, examining the incidence of disks around young stellar objects through
their infrared excess, and studying the galactic ecology through extended 3.3~\mm\ PAHs
emission in star forming complexes. Despite its modest size, it obtained the deepest
image in the 3.5~\mm\ L-band then obtained (in 1999), of the embedded stellar content in
the 30 Doradus region of the LMC \citep{Maercker_Burton_05}, detecting a source as faint
as $\mab=14.5$. This was only bettered by the 8~m VLT in 2004.

Additionally, the International Robotic Antarctic Infrared Telescope
\citep[IRAIT;][]{Tosti_e_06}, which is currently being deployed to Dome~C, should be
operating over the next few winter seasons. IRAIT is a 0.8~m infrared telescope equipped
with a double-arm camera, AMICA, capable of imaging in a series of broad- and narrow-band filters over the range 2--28 \mm\ with a field-of-view of $\sim2.5'\times2.5'$ in each
arm \citep{Dolci_e_06}. The primary scientific motivation for IRAIT \citep{Guandalini_e_08} is to conduct wide-area surveys of dense ISM regions and star forming regions, to search for intrinsically cool objects such as Brown Dwarfs, and to perform wide-area surveys to investigate mass-losing evolved (Asymptotic Giant Branch) stars. It is expected that IRAIT will provide valuable information relevant to many aspects of the PILOT project, including technological solutions to the problems arising from the low-temperature and high relative humidity environment; observatory operating, control, and communications requirements; science case development and planning; and observing strategies.

\subsection{Unexpected Science Outcomes}
For the majority of astronomical facilities, the initial science case for that facility
has not predicted many of the key scientific outputs. Such new and unexpected scientific
discoveries occur, in general, whenever technological, engineering or site-related
advances lead to large increases in the observational parameter space relative to
existing facilities. There are many examples of such unexpected science outcomes
throughout the history of astronomy. The National Geographic Palomar Observatory Sky
Survey conducted during the 1950s led to the discovery, for example, of the rich
properties of galaxy clusters by \citet{Abell_58}. The 3CR catalogue of bright radio
sources published by \citet{Bennet_62} led to the quasar discovery of \citet{Schmidt_63}
and \citet{Matthews_Sandage_63}. The Hubble Deep Field \citep{Williams_e_96} was not part
of the original science case for the HST, yet has been one of its most important
outcomes. The Sloan Digital Sky Survey resulted in a wealth of science not originally
anticipated, including the discovery of Galaxy halo streams \citep{Yanny_e_03}, thousands
of cataclysmic variables \citep{Szkody_e_06}, the solar abundances of QSO absorption
lines \citep{York_e_06a}, the strict mass metallicity relationship for galaxies
\citep{Bernardi_e_05}, the faint end slope of the quasar luminosity function
\citep{Richards_e_05}, and the colour segregation of hundreds of thousands of asteroids
\citep{Szabo_e_04}. While the Spitzer science case\footnote{NASA JPL, 1997,
\url{http://ssc.spitzer.caltech.edu/documents/SRD.pdf.}} proposed searches for exoplanets
via direct imaging, it did not predict the possibility, as recently demonstrated (e.g.,
\citep{Knutson_e_08}, of characterising exoplanet atmospheres via primary transit water
vapour absorption or the secondary eclipse methods.

As with these historical examples, it is likely that some of the most exciting results to
come from PILOT are not included in this science case. We can be reasonably confident of
this since the unique atmospheric characteristics of the Dome~C site, combined with a
specific instrument suite designed to take advantage of these characteristics, opens a
new discovery space.

\section{Observing Strategies}
\subsection{Shared Cadence}
Several of the proposed science programs for PILOT require dedicated observing time in
targeted fields, e.g., gamma-ray burst follow-up from satellite alert (Paper II: Section
2.2), imaging of galaxy clusters (Paper II: Section 3.2), time-series observations of
\ocen\ (Paper III: Section 2.4), and exoplanet follow-up observations (Paper III: Section 4.3). There are many other projects, however, that have overlapping observational
requirements. Adopting a shared-cadence approach to these projects, i.e., by aiming to
generate a dataset that can be used to satisfy the requirements for a range of science
objectives, is likely to significantly increase the scientific efficiency of the PILOT
facility. This strategy is much better suited to Dome~C operation than more traditional
observing strategies with multiple projects and observing sequences that change on a
nightly basis.

Three separate projects identified here require deep observations in near-infrared bands
over wide areas of the sky: the search for pair-instability supernovae at very high
redshift (Paper II: Section 2.1), the search for high-redshift galaxies (Paper II:
Section 3.1), and the search for dusty SN at moderate redshift (Paper II: Section 4.2).
While the most appropriate target fields still need to be determined for each of these
projects, it is likely that there will be some overlap in the preferred fields, which
will be well away from the Galactic plane. These projects can thus be pursued with a
shared cadence. This would necessarily involve some compromise on observing strategy, as
the PISN search requires repeat observations with periods up to several hundred days, the
SN Ia search requires repeat observations with a period in the range 4--14 days, and the
galaxy survey does not require any repeat observations. Such compromises are likely to
lower the detection rate per observing time for each project, but to increase the overall
efficiency when combined.

Similarly, the star and planet formation studies (Paper III: Section 3) each have
observational requirements for multi-band imaging of star forming regions throughout the
mid-infrared. In particular, the Chamealeon dark clouds complex has been identified as a
suitable site to search for circumstellar discs around young stellar objects and brown
dwarfs (Paper III: Section 3.2), crystalline silicate signatures around similar objects
(Paper III: Section 3.3), and embedded class 0/I young stellar objects (Paper III:
Section 3.4). While each of these projects requires a distinct set of observing bands
(and spectral resolution) there is potential for a shared observing strategy in regions
such as the Chamealeon dark clouds complex. There may also be some overlap with the
detailed H$_{2}$ maps produced as part of the proposed Galaxy ecology project (Paper III:
Section 3.1).

Additionally, it should be expected that all of the proposed large-area survey projects
will contain scientific ``by-products" in their datasets; the number and significance of
these by-products will be strongly dependent on the observing strategy adopted. The
weak-lensing survey (Paper II: Section 4.1) in the visible, for example, will image a
large region of sky over several seasons. While it will require multiple exposures for
each field to reduce systematic errors in shear measurement, there is no specific
requirement on time sampling. The required depth per field could be built up by sampling
each field with some time delay. With a multi-epoch approach as proposed, for example,
for Skymapper \citep{Keller_e_07} and LSST \citep{Ivezic_e_08}, the discovery and
characterisation of a large number of variable and transient objects (supernovae, solar
systems objects, long- and short-period variable stars, quasars, exoplanet transits, etc.) with periods ranging from hours to years would be possible. Adopting such a strategy
would necessarily increase the time required to complete the full weak-lensing survey. To decide on the most appropriate observing scheme it is thus important (for all survey
projects) to identify the additional scientific output (in comparison with other proposed facilities) and to determine how this output is effected by the observing cadence chosen
and hence the total observing time required.

\subsection{Dynamic Scheduling}
In combination with a shared-cadence strategy, the observational efficiency and hence
scientific output of the PILOT facility would be significantly increased by employing a
dynamic-scheduling approach to observing. As discussed in Section 2, there are a variety
of advantages that the Dome~C site conditions offer, such as high thermal infrared
sensitivity, high photometric accuracy, and good image quality. Each science case targets these advantages in different ways and thus makes specific demands on the required
observing conditions.

The efficiency of the weak-lensing survey is a strong function of the obtained imaging
FWHM (full width half maximum). It is known that there are extended periods of
exceptional seeing at Dome~C that correspond either to calm conditions in the free
atmosphere or a reduction of the turbulent boundary layer height to below the height of
the telescope (E. Fossat, private communication). Doing this survey in only the best 50\% of conditions (in terms of resolution) would be a more efficient use of facility time, than dedicating 100\% of time over a full winter season. Similarly, in order to achieve the potential diffraction-limited performance with the Lucky Imaging camera, not only is the integrated seeing important, but also the atmospheric coherence time and isoplanatic
angle. These factors are usually, but not necessarily, correlated with the seeing.

\begin{figure*}[t]
\begin{center}
\includegraphics[width=15cm]{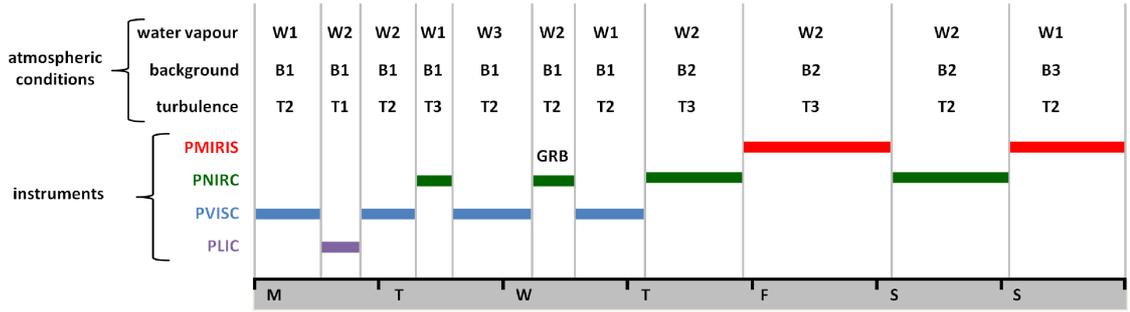}
\caption{Illustrative observing schedule for PILOT over 1 week during mid-winter.
Instruments are dynamically selected based on the atmospheric characteristics
(water vapour, sky background, and turbulence conditions), and science objectives. Atmospheric conditions are ranked into a range of bands, e.g., T1 = 10\% best seeing, B1 = dark sky, etc., according to the real-time turbulent statistics, the atmospheric temperature and solar/lunar elevation, and the atmospheric precipitable water vapour content. Several science projects can override atmospheric scheduling (e.g., GRB and planet alerts).}\label{week}
\end{center}
\end{figure*}

For optical survey projects, an efficient use of sky background conditions is essential.
For example, the proposed weak-lensing survey requires dark skies, whereas the
time-series observations of open clusters can be achieved in twilight conditions.

Survey programs in the near-infrared, i.e., with PNIRC, depend only weakly on the seeing
(because in median conditions the resolution is close to diffraction limited) but are
strongly dependent on atmospheric thermal emission (for wavelengths above \si2.4~\mm) and solar zenith-angle (for wavelengths below \si4~\mm).

Observations with the PMIRIS camera, i.e., above \si7~\mm, are essentially diffraction
limited regardless of the turbulence conditions, and are insensitive to the solar zenith
angle. To obtain high sensitivity at these wavelengths requires low atmospheric thermal
emission. Projects requiring narrow-band observations at wavelengths corresponding to
atmospheric water vapour absorption, such as proposed for the H$_{2}$ mapping project,
are preferentially done with low precipitable water vapour conditions. For some projects
(e.g., exoplanet secondary transits) the stability of the thermal atmospheric emission is more important than water vapour.

To satisfy these disparate demands on observing conditions will require a real-time
nested scheduling strategy, where both instruments and science projects are prioritised
and attributed to appropriate observing conditions. Such an observing strategy is
illustrated in Figure~\ref{week} for a one week period during mid-winter. This approach
adds complexity to the overall facility. Physically, a suite of real-time site monitoring instruments will be required and all instruments need to be relatively easy to
interchange both remotely and automatically. There will also be additional software
overheads associated with such an approach. This increase in complexity (hence cost) is
far outweighed, however, by the benefits in terms of total scientific output and facility efficiency.

\subsection{A Long-Term Science Program}

Is it intended that there will be several large ``flagship" science projects for the
PILOT facility, and that the majority of observing time will be dedicated to these
projects. These projects will include the four primary science mission projects described in Section 3.3. A number of other science projects will also be selected as priority
science. Additionally, it is scientifically beneficial that some fraction of telescope time be devoted to smaller-scale ``queued" projects that are proposed throughout the lifetime of the facility.

\begin{figure*}[t!]
\begin{center}
\includegraphics[width=13cm]{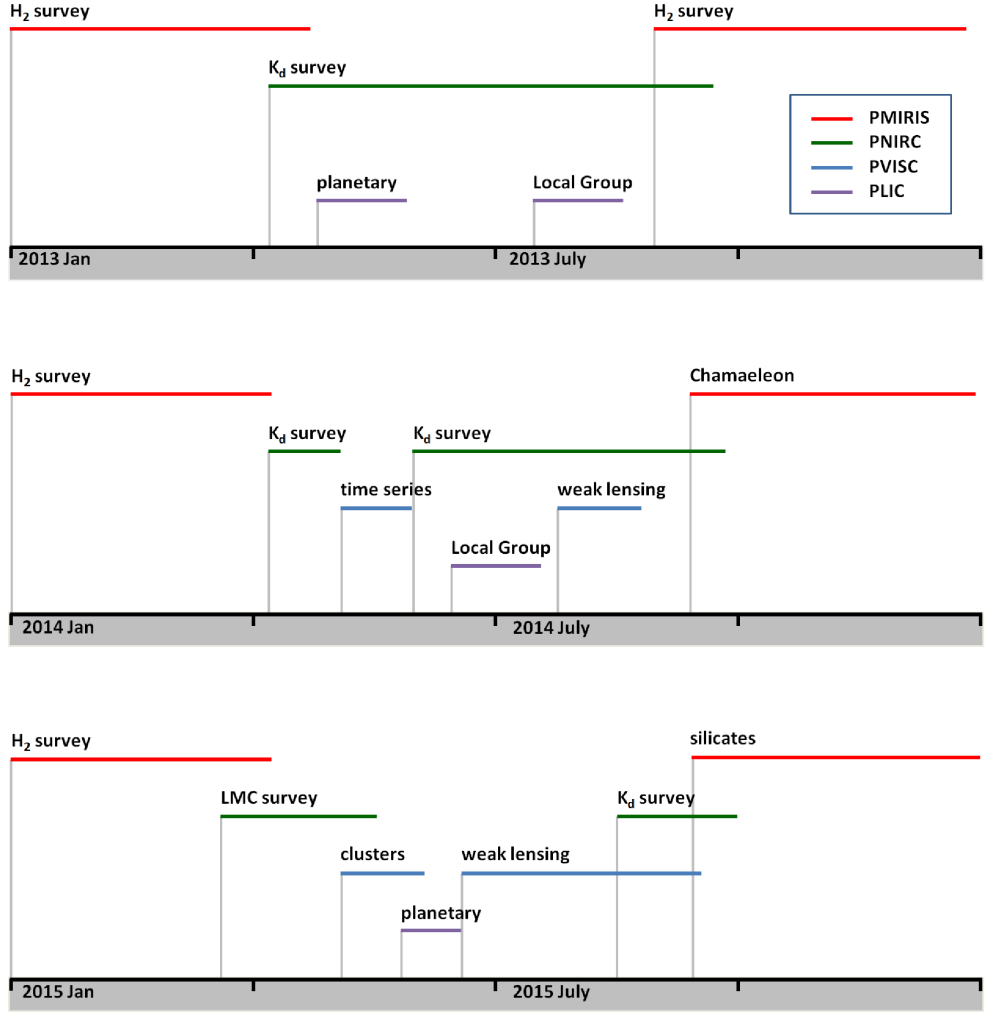}
\caption{Illustrative observing schedule over a 3~year period for PILOT, showing the
primary science project and instrument for each period. Overlapping regions represent
dynamic scheduling that occurs on a much finer time-scale. It is likely that the first
commissioning season (not shown here) will include extensive operation of the PLIC
instrument. Here we assume that the PNIRC and PMIRIS instruments are deployed together,
and that the PVISC camera is installed the following year. This illustrative scheme does
not account for the time of year that objects are in appropriate locations for
observation.}\label{years}
\end{center}
\end{figure*}

While several options have been considered, the mechanism for time allocation of PILOT
science projects has not yet been determined. The processes by which the science cases
are prioritised, the telescope access policy, and the fraction of time available to
primary, secondary, and queued projects, will be determined in the next design phase for
PILOT. All of the science cases will then be further developed and refined before
undergoing a selection process that will allow the observing strategy to be finalised and a motivated long-term scientific program to be developed.

During the next design phase for the PILOT project, a strategy for instrument deployment
will also be determined. It is likely that a phased implementation of instruments will be employed based on a sensitivity matrix analysis of the telescope scientific
functionality. One possible scenario is that PLIC will be the first instrument deployed,
as it is simpler and less expensive than the other baseline instruments and it will allow a detailed characterisation of the site and telescope performance during its
commissioning phase.

In Paper II and Paper III, the observational requirements for a broad range of PILOT
science projects are presented. Although many of these projects require large amounts of
observing time, the majority of science projects presented could be accomplished over the ten year lifetime of the PILOT facility, as illustrated in Table~\ref{Table3}.

In Figure~\ref{years}, we present an illustrative PILOT observing program for a three
year period. Observations with PMIRIS are possible in the summer months at Dome~C. As
shown, it should be possible to complete a molecular hydrogen survey for a wide area of
the Galactic Plane in two complete observing seasons. This would allow summer-time
observations in the following years to be devoted to other mid-infrared projects, such as surveys of the Chamaeleon dark clouds complex. By adopting a shared-cadence strategy, a
large fraction of the proposed \kd\ projects (pair-instability supernovae searches, Type
Ia supernovae searches, and high-redshift galaxy surveys) could be completed in a three
year period. Other infrared survey projects (e.g., of the Magellanic Clouds) could then
be undertaken in the following years. The weak-lensing optical project is likely to take
at least five years to complete. Other smaller-scale optical projects requiring either
twilight conditions (e.g., open cluster asteroseismology) or dark sky conditions (e.g.,
galaxy clusters) could be completed well before then if allocated some fraction of good
quality observing conditions each year. It is likely that with a full year of
commissioning, the majority of PLIC projects could be completed in a few years.

\section{Synergies with other Facilities}

The role of other facilities in contributing to PILOT science is
addressed in detail in each of the science projects in Papers II and
III. Here, we consider the inverse question: what value can PILOT add
to the primary science from other telescopes? In particular, we
consider the Giant Magellan Telescope (GMT), the South Pole Telescope
(SPT), the Australian Square Kilometre Array Pathfinder (ASKAP), and
the Murchison Widefield Array (MWA). This list is meant to be
illustrative rather than exhaustive. For example, other large
next-generation optical telescopes could be substituted for GMT.

\subsection{Giant Magellan Telescope}
The Giant Magellan Telescope is proposed as a 25~m diameter optical
and infrared telescope to be sited at Las Campanas Observatory in
Chile. The telescope is composed of seven 8.4~m primary mirrors and
has an effective collecting area of 21.4~m. Proposed instruments
include an optical high-resolution spectrometer, a near-infrared high
spectral-resolution spectrometer, and a mid-infrared imaging
spectrometer.

\begin{figure}[t]
\begin{center}
\includegraphics[width=7.8cm]{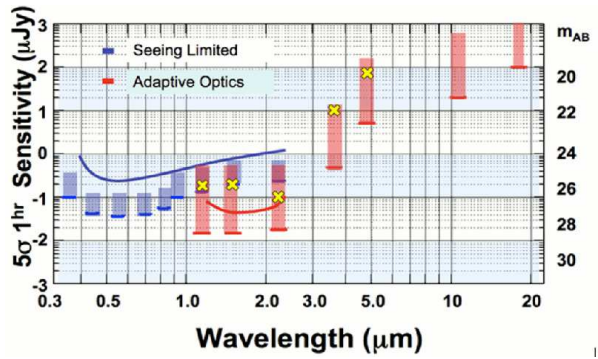}
\caption{Sensitivity limits (for a 5~$\sigma$, 1~hour observation) for GMT. The curves
show the limits for intermediate resolution spectroscopy. The shaded vertical bars show
the gains over current 8~m telescopes. The top of the shaded bars are the 8~m
sensitivity, the bottom is the GMT sensitivity. From the \emph{Giant Magellan Telescope
Science Case}. The yellow crosses show the wide-band imaging sensitivity for PILOT with
PNIRC (for a 5~$\sigma$, 4~hour observation). The PNIRC field-of-view is more than 100
times the area of the diffraction-limited GMT field-of-view. }\label{GMT}
\end{center}
\end{figure}

The GMT science case\footnote{\url{http://www.gmto.org/sciencecase}}
has a wide range of science drivers, many of while could benefit from
PILOT observations. The GMT high resolution near- and mid-infrared
sensitivities are well matched to the PILOT wide-field wide-bandwidth
sensitivities, as illustrated in Figure~\ref{GMT}.

Since the first-light for the GMT will likely occur after PILOT has
achieved its primary science goals, the various PILOT wide-area
surveys will be immediately useful to search for GMT infrared sources.  For
example, the PILOT \kdb\ surveys, in conjuction with photometric
redshifts from Spitzer, would form an ideal high-redshift (out to $z
\approx 7$) galaxy source catalogue for GMT. GMT could obtain high
spatial and spectral resolution ($R = 3000$--100,000) of the galaxies
and use them for several key projects including the measurement of the
evolution of dark energy through the exploration of baryonic
oscillations, and the examination of the epoch of reionisation through
the Lyman-$\alpha$ continuum break in high-redshift quasars and
galaxies. In the mid-infrared, PILOT's wide-area surveys will pinpoint
interesting circumstellar discs for follow-up with the GMT at high
spatial and spectral resolution.

After PILOT's primary science mission is completed, PILOT's
main role could be as a ``finderscope'' for telescopes such as GMT.

\subsection{South Pole Telescope}

The South Pole Telescope is a 10~m diameter millimetre wave telescope
located at the US Amundsen-Scott South Pole station. The SPT achieved
first-light in 2007 and provides a 1~\degsq\ field-of-view with
\si1$'$ resolution in the frequency range 95--345~GHz. The SPT makes
use of the exceptionally low precipitable water vapour in the
atmosphere above the South Pole, as well as its high temporal stability
\citep{Chamberlin_01}.

The primary science goal for the SPT is to measure the redshift
evolution of the abundance of massive galaxy clusters in order to
constrain cosmological parameters such as dark energy. It makes use of
the Sunyaev-Zeldovich effect (SZE) in which Cosmic microwave
Background (CMB) radiation is scattered by the gas in distant galaxy
clusters \citep[see, e.g.,][]{Carlstrom_e_02}. The magnitude of the
SZE is directly related to the cluster mass but independent of
redshift, making it an ideal cosmological probe. With a survey of
\si4000~\degsq, some 20\,000 galaxy clusters are expected in the
redshift range $z \approx 0.1$--2 \citep{Ruhl_e_04}. Optical and
near-infrared follow-up observations will be required for photometric
redshift determination of this large sample of clusters. For objects out to $z = 1.3$, redshifts will be obtained by the (\emph{g}, \emph{r}, \emph{i}, \emph{z}) Dark Energy Survey (DES) on the Blanco 4~m telescope over a period of 5~years \citep{Carlstrom_e_06}.

PILOT could provide an independent estimate of the masses of the SPT
galaxy clusters through strong lensing events identified from
surveys. The weak-lensing survey will obtain optical images of
galaxies out to a redshift of $z\approx1$ over a wide region of the
sky. Such a survey would be expected to find numerous nearby galaxy
clusters that are strongly lensing more distant background
galaxies. Similarly, the proposed PILOT near-infrared surveys would
find strongly-lensed galaxy-clusters at higher redshifts. The cluster
masses determined from the PILOT data would be a useful verification
of the SPT SZE results and a valuable test for many potential
systematic effects. Additionally, PILOT near-infrared survey data
could provide photometric redshifts for SZE galaxy clusters at a
higher redshift ($z = 1.3$--2.0) than possible with the DES optical
data.

\subsection{Australian Square Kilometre Array Pathfinder}

The Australian Square Kilometre Array Pathfinder (ASKAP) is a next
generation radio telescope acting as a pathfinder for the future Square
Kilometre Array (SKA). ASKAP consists of an array of
30--45 dishes of 12~m diameter observing a field-of-view of
30~\degsq\ in the frequency range 700--1800~MHz. ASKAP is planned for deployment and commissioning in 2012. It will thus be operational throughout the lifetime of PILOT.

Key science goals for ASKAP \citep{Johnston_e_07} include the detection of a statistically significant sample of low redshift ($z = 0.05$--0.2) galaxies
via a H\textsc{I} 21~cm survey; a confusion limited all-sky
1.4~GHz continuum survey to detect on the order of 60 million
galaxies; and the detection of a large number of radio transient
events. In several of these areas, PILOT can provide a complementary role.

Although the vast majority of galaxies detected by ASKAP will be
normal (Milky Way type) galaxies at low redshift ($z<0.3$), a number
of objects, such as radio galaxies (AGN) and extreme starbursts at
redshifts out to $z = 6$ will also be detected. Near-infrared survey
data, i.e., from the Vista Hemisphere Survey (VHS) project, can be
used to determine the redshift of these galaxies via the $K-z$
relationship \citep{Jarvis_e_01}.  The depth of the VHS survey is
$\mab\approx22$ at \kb, which corresponds to a redshift of
$z\approx4$. Although there is some question about the validity of the
$K-z$ relationship for high redshift \citep{De_Breuck_e_06}, objects
in the ASKAP galaxy sample that are not detected by other surveys
could potentially be identified with deep \kd\ PILOT imaging using
PNIRC.

Additionally, by virtue of its extreme southerly
latitude as well as its similar longitude, PILOT is uniquely placed to
rapidly follow-up any ASKAP transient event in a range of
wavelengths (from the visible to the mid-infrared). Analysis of archival VLA data \citep{Bower_e_07} suggests that a wide-field radio survey will find a large number of transients, many of which will be of unknown origin. Immediate searches for
optical/infrared counterparts to the radio emission, and later
follow-up searches for host systems, will be important steps to
determine the nature of many of these objects.

\subsection{Murchison Widefield Array}
The MWA is a low frequency radio array that will be sited in the
Murchison Radio Observatory in Western Australia. The array consists
of 8000 dual-polarization dipole antennas operating in the 80--300~MHz
range. The majority of the tiles (each a $4 \times4$ array of dipoles)
will be concentrated within a \si1.5~km core region, allowing a
resolution of a few arcmins over a field-of-view of \si200~\degsq. The
remaining tiles will be located outside this core region allowing
higher angular resolution for some measurements. The MWA should be
completed by the time that PILOT achieves first light.

There are three key science goals for the MWA: probing the epoch of
reionisation via sensitive observations of emission and absorption in
the 21~cm hyperfine transition line of neutral hydrogen that are
redshifted below \si200~MHz \citep{Morales_Hewitt_04};
solar-heliospheric-ionospheric studies of solar wind parameters,
crucial to the coronal mass ejections \citep{Salah_e_05}; and the
search for low frequency transients occurring from gamma-ray bursts,
compact objects, and stellar and planetary emission.

\begin{figure}[t]
\begin{center}
\includegraphics[width=7.0cm]{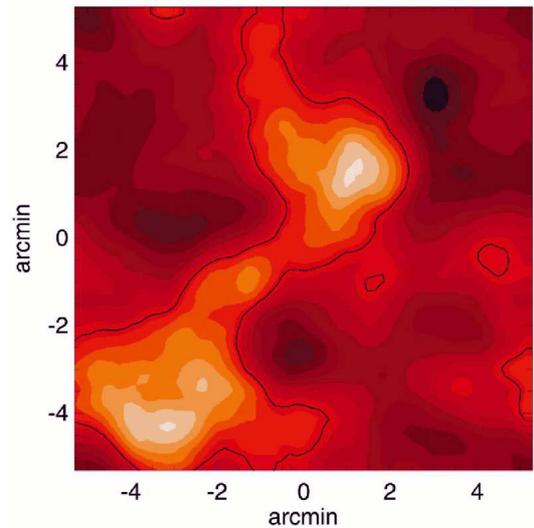}
\caption{Simulation of redshifted 21~cm emission/absorption at
  $z\approx8.5$. From \citet{Tozzi_e_00}. The MWA aims to characterize
  such structure. The PILOT \kd\ galaxy survey would be expected to
  find at least four star forming galaxies, and a larger number of
  evolved galaxies at a redshift in the range $z=5$--7 in a field this
  size.  }\label{MWA}
\end{center}
\end{figure}

Each of the main science drivers for MWA will benefit from PILOT
observations. The MWA observations of the cosmic neutral hydrogen over
wide fields at high redshifts can be cross-correlated against PILOT
measurements of the oldest and brightest distant galaxies, as
illustrated in Figure~\ref{MWA}. The MWA observations of solar wind
and coronal mass ejection can be compared to lower wavelength solar
observations with the proposed SmilePILOT sub-millimetre
camera. Similarly to the case for ASKAP, PILOT can be used to
follow-up a subset of MWA low-frequency transient sources at optical
and infrared wavelengths.

\section{The Future of Antarctic Astronomy}

Despite its moderate sized aperture, PILOT is a powerful facility that is suited to many
specific roles where it fills a favourable or even unique position in performance space,
as illustrated in Section 2.6. This is also demonstrated by the significance and breadth
of the scientific projects proposed in this series of papers. PILOT, however, is intended as the first step towards a large-scale astronomical observatory at Dome~C. We therefore
discuss here what the future may hold.

While an ELT scale telescope at Dome~C, as proposed by \citet{Angel_e_04}, would be
exceptionally powerful for a wide range of science, it is not likely to be built within
the next 20~years. Here we concentrate on the facility to follow PILOT on a shorter
timescale. A range of projects have been proposed that take advantage of the Dome~C
conditions to achieve a spatial resolution (using high-order or ground-layer adaptive
optics systems) that is 2--3 times higher, a near-infrared sensitivity that is 2
magnitudes deeper, and a photometric precision that is 3--4 times more accurate, than any temperate observatory. Options for this next-generation Dome~C facility, that could conceivably be operational in the time-frame of the next 10--15 years, include:

\begin{itemize}
\item LAPCAT, Large Antarctic Plateau Clear Aperture Telescope: an 8.4~m diameter
    off-axis telescope operating in the optical and infrared, with a highly efficient adaptive optic system \citep{Lawrence_e_08} optimised for high-resolution imaging of exoplanets \citep{Storey_e_06};
\item IRAST, InfraRed Antarctic Survey Telescope: an 8--10~m class infrared telescope with a ground layer adaptive optics system operating over a very wide field-of-view, dedicated to deep infrared survey projects that probe the first galaxies and stellar populations to form;
\item ASO, Antarctic Sub-millimetre Observatory: a 12~m telescope operating at
    far-infrared and sub-millimetre wavelengths, with a range of science goals
    including the search for high-redshift optically-obscured starburst galaxies and
    the molecular characterisation of the interstellar medium \citep{Olmi_e_05};
\item Xian: a 400 element array of 0.5~m wide-field (\si20~\degsq) Schmidt telescopes which are observing a large fraction of the sky simultaneously in order to search    for optical transients \citep{York_e_06b};
\item ALADDIN, Antarctica L-band Astrophysics Discovery Demonstrator for Interferometric Nulling: an infrared interferometer consisting of two 1-metre diameter telescopes mounted on a 40-metre diameter circular structure, for the characterisation of the zodiacal light for future space interferometer missions \citep{Coude_e_05}.
\item KEOPS, Kiloparsec Explorer for Optical Planet Search: a 36 element
    interferometric array of 1.5~m off-axis telescopes, dedicated to the detection
    and characterisation of exoplanets via infrared nulling \citep{Vakili_e_05};
\item WHAT: a LAMOST style, 8~m diameter aperture, wide-field, fixed-axis,
    all-reflective Schmidt telescope operating in the optical and near-infrared,
    suited to a range of science from galaxy evolution to large scale structure to
    stellar population studies of nearby galaxies \citep{Saunders_McGrath_03}.
\end{itemize}

\section{Conclusion}

The advantageous atmospheric and ground-level site conditions found at Dome~C on the
Antarctic plateau, combined with an optimised optical configuration and instrumentation
suite designed for these conditions, allow PILOT to compete with or outperform
larger-aperture telescopes located elsewhere, over a range of wavelengths. In the
visible, the wide-field mode for PILOT has a spatial resolution that is 2--3 times higher than other existing or planned ground-based telescopes, and a survey speed that is faster than any existing ground- or space-based facility. In high-resolution mode, using a Lucky Imaging system at visible wavelengths, PILOT should achieve a spatial resolution
approaching the diffraction limit. In the near-infrared, PILOT has a survey speed that is substantially faster than any existing telescope. In the mid-infrared, it is the only
telescope capable of obtaining moderate spatial resolution photometry over very wide
regions of sky.

A series of science projects have been identified that take advantage of the unique
performance space for this telescope. These projects, which have been summarised in this
paper, are described in more detail in the two companion papers in this series. The
questions that can be addressed with PILOT cover a wide range of astronomical topics
examining the Universe at a range of distance scales: from studies of the the first stars that form at very high redshift, to studies of the large-scale structure of the Universe
at moderate redshift, to studies of nearby galaxies and stellar clusters, to studies of
star and planet formation in the Milky Way, to studies of the Sun and planets within the
Solar System. The discovery space for PILOT will also allow it to complement the
scientific objectives of a variety of other existing and planned facilities operating
over a broad spectral range: at visible and infrared wavelengths (e.g., GMT), at
millimetre wavelengths (e.g., SPT), at radio frequencies (e.g., ASKAP), and at low
frequencies (e.g., MWA).

A second objective for the PILOT project is to act as a pathfinder for future, more
ambitious Antarctic optical/infrared telescopes. In this role, PILOT must demonstrate
that the Dome~C site conditions can be fully utilised, and prove the technical and
logistical feasibility of operating a large optical/infrared telescope in the Antarctic
plateau environment. It must be recognised that to take full advantage of the Dome~C site conditions will require a larger aperture 8--10 m class telescope; such a facility would be an exceptional tool for astronomy.

\section*{Acknowledgments}
The PILOT Science Case, presented here, was produced as part of the PILOT conceptual
design study, funded through the Australian Department of Education, Science, and
Training through the National Collaborative Research Infrastructure Strategy (NCRIS)
scheme, and the University of New South Wales through the UNSW PILOT Science Office. The
European contribution has been supported by the ARENA network of the European Commission
FP6 under contract RICA26150. We thank Eric Fossat and Giles Durand for providing us with
data prior to publication.


\end{document}